\documentclass[a4paper,12pt]{article}
\usepackage{amssymb,amsmath}
\usepackage{a4wide}
\usepackage{epsfig}

\newcommand{\RR}{{\mathbb{R}}}

\newcommand{\ZZ}{{\mathbb{Z}}}
\newcommand{\pa}{\partial}
\newcommand{\ii}{i}
\newcommand{\ee}{{\mathrm{e}}}
\newcommand{\zb}{\bar{z}}
\newcommand{\Kb}{\overline{K}}

\newcommand{\quar}{\tfrac{1}{4}}
\newcommand{\tr}{\mathop{\mathrm{tr}}\nolimits}

\newcommand{\cM}{{\cal M}}

\newcommand{\cS}{{\cal S}}

\newcommand{\cR}{{\cal R}}

\title{Geometry of Solutions of Hitchin Equations on $\RR^2$}
\author{R.\ S.\ Ward\footnote{email address: richard.ward@durham.ac.uk}
  \bigskip
  \\Department of Mathematical Sciences,
  \\Durham University, Durham DH1 3LE.}
\date{\today}

\begin{document}

\maketitle

\begin{abstract}
\noindent We study smooth SU(2) solutions of the Hitchin equations on
$\RR^2$, with the determinant of the complex Higgs field being a
polynomial of degree $n$. When $n\geq3$, there are moduli spaces
of solutions, in the sense that the natural $L^2$ metric is well-defined
on a subset of the parameter space. We examine rotationally-symmetric
solutions for $n=1$ and $n=2$, and then focus on the $n=3$ case,
elucidating the moduli and describing the asymptotic geometry as well as the
geometry of two totally-geodesic surfaces.
\end{abstract}

\newpage

\section{Introduction}
The dimensional reduction of the self-dual Yang-Mills equations to
$\RR^2$ has long been of interest, even though this system does
not admit smooth finite-energy solutions with compact gauge group
\cite{L77, S84}. But there are solutions if one allows singularities
and/or infinite energy \cite{S81, S84}, or for complex or non-compact
gauge group \cite{B84, MJ07}. The equations are conformally invariant,
and so one can also study the system on general Riemann surfaces;
in that context, the equations are known as the Hitchin equations,
and in particular there are smooth solutions on compact
Riemann surfaces of genus at least two \cite{H87}. The $\RR^2$
case, with appropriate boundary conditions, may be viewed as a system
on $S^2$ with a singularity at infinity; and more generally one could allow
several designated singularities.
There is a natural $L^2$ metric on the space of solutions, although the
integral defining it may not always converge; these moduli-space metrics,
when well-defined, are hyperk\"ahler \cite{H87, BB04}. Such Hitchin systems and
their moduli spaces have close connections with supersymmetric field theory:
see, for example, reference \cite{N14}.

The focus in this paper is on smooth SU(2) solutions of the Hitchin equations
on $\RR^2$. There is a natural boundary condition which involves a positive
integer $n$: namely, the determinant $H$ of the complex Higgs field
$\Phi$ is taken to be a polynomial of degree~$n$ in $z=x+\ii y$. Then the
solutions generically resemble $n$ lumps on $\RR^2$, located at the
zeros of $H$: the solutions are parametrized by these locations, together
with relative angles between the lumps. Some of these parameters have
$L^2$ variation, and these are moduli; this is only possible when $n\geq3$,
and the main aim below is to describe some of the geometry of the $n=3$
moduli space $\cM$. In particular, we calculate the asymptotic geometry by
using a singular approximation to the fields, and obtain the metric on
two natural totally-geodesic surfaces in $\cM$ with the help of a numerical
calculation. To begin with, however, we recall the rotationally-symmetric
$n=1$ solution, and describe a one-parameter family of
rotationally-symmetric $n=2$ solutions.

Some of the methods used below first arose in studies of the Hitchin equations
on the cylinder $\RR\times S^1$, in connection with periodic monopoles
\cite{CK01, CK02, CK03, HW09, MW13}. Periodic monopoles correspond,
via the generalized
Nahm transform, to solutions of the Hitchin equations on the cylinder
$\RR\times S^1$ satisfying appropriate boundary conditions.
Since the cylinder is conformally equivalent to the punctured plane,
these Hitchin fields are defined on $\RR^2$ minus a point
(or $S^2$ minus two points), and so they do not
overlap with the solutions considered below.


\section{The Hitchin equations on $\RR^2$}
Let $A_j$ be a smooth SU(2) gauge potential on $\RR^2$, with corresponding
gauge field $F:=F_{xy}=\pa_x A_y-\pa_y A_x+[A_x,A_y]$, and let $\Phi$
denote a Higgs field taking values in the complexification of the Lie algebra
${\mathfrak su}(2)$. The Hitchin equations \cite{L77, H87} are
\begin{equation}  \label{H1}
  D_{\zb}\Phi = 0, \quad F = \frac{\ii}{2}\, [\Phi,\Phi^*].
\end{equation}
Here $z=x+\ii y$ is a complex coordinate on $\RR^2$,
$D_{\zb}\Phi:=\pa_{\zb}\Phi+[A_{\zb},\Phi]$ is the covariant derivative,
and $\Phi^*$ denotes the complex-conjugate transpose of $\Phi$.
The system (\ref{H1}) is completely-integrable in the sense of having
a Lax pair, namely the reduction of the Lax pair for the self-dual
Yang-Mills equations \cite{MW96}. In addition, for example, the moduli
spaces of solutions on Riemann surfaces may be regarded as
completely-integrable Hamiltonian systems \cite{H87}.

It follows from (\ref{H1}) that the determinant $\det\Phi$ is holomorphic in $z$;
as a boundary condition, we require that $H(z)=\det\Phi$ should be a
polynomial in $z$, and that $F\to0$ as $|z|\to\infty$. Such solutions generically
have $F$ peaked at the zeros of $H(z)$, and we may visualize these as lumps.

Suppose now that we have a family of solutions depending on a parameter $t$.
Let $\dot{\Phi}$ and $\dot{A}_{\zb}$ denote the $t$-derivatives of
$\Phi$ and $A_{\zb}$, representing a vector $V$ on the space
of solutions. If we impose a condition that this variation $V$ be orthogonal
to the gauge orbit at $(\Phi, A_{\zb})$, then the combined equations
satisfied by $(\dot{\Phi},\dot{A}_{\zb})$ are
\begin{equation}  \label{PertEqns}
  D_{\zb}\dot{\Phi}=[\Phi,\dot{A}_{\zb}], \quad 
             4D_{\zb}\dot{A}_{z}=[\Phi,\dot{\Phi}^*].
\end{equation}
Given that  $(\dot{\Phi},\dot{A}_{\zb})$ satisfy (\ref{PertEqns}),
the natural $L^2$ norm of $V$ is
\begin{equation}  \label{L2norm}
   ||V||^2=\frac{1}{2}\int \tr\left(\dot{\Phi}\dot{\Phi}^*+
                    4\dot{A}_{\zb}\dot{A}^*_{\zb}\right)\,|dz|^2.
\end{equation}
Although (\ref{L2norm}) is not the only metric one can define on a
solution space or moduli space, it is the most natural one both from
the geometrical point of view \cite{H87} and in relation to the dynamics
of topological solitons \cite{MS04}.
If (\ref{L2norm}) converges, then $t$ is a modulus; whereas if (\ref{L2norm})
diverges, we refer to $t$ as a parameter. The moduli space $\cM$ is the space
of moduli, for fixed values of the parameters.

A useful choice of gauge was described in \cite{HW09, GMN13}; it may be
summarized as follows. The gauge field $F$ is taken to be in the
$\sigma_3$-direction in ${\mathfrak su}(2)$, and $\Phi$ has the form
\begin{equation}  \label{psi-form}
  \Phi=\left(\begin{matrix} 0 & \mu_+\ee^{\psi/2} \\ 
    \mu_-\ee^{-\psi/2} & 0\end{matrix}\right).
\end{equation}
Here $\mu_{\pm}$ are polynomials in $z$ satisfying $\mu_+ \mu_-=-H$,
and $\psi$ is a smooth real-valued function. The gauge potential has the form
\begin{equation}  \label{A-form}
A_{\zb}=-\quar(\pa_{\zb}\psi)\,\sigma_3 + \alpha\Phi,
\end{equation}
where $\alpha$ is smooth complex-valued function. The Hitchin
equations then reduce to equations on $\psi$ and $\alpha$, namely
\begin{eqnarray}
\Delta\psi &=& 2(1+4|\alpha|^2)(|\mu_+|^2\ee^{\psi}-|\mu_-|^2\ee^{-\psi}),
                         \label{psi-eqn} \\
0 &=& \ee^{-\psi/2}\pa_{z}(\ee^{\psi}\mu_+\alpha)+
         \ee^{\psi/2}\pa_{\zb}(\ee^{-\psi} \bar\mu_+ \bar\alpha), \label{alpha-eqn}
\end{eqnarray}
where $\Delta=\pa_x^2+\pa_y^2=4\pa_{z}\pa_{\zb}$.
The residual gauge freedom consists of constant diagonal SU(2)
transformations, plus conjugating by $\sigma_1$. In the special case
$\alpha=0$, the equation (\ref{psi-eqn}) is locally equivalent to the
sinh-Gordon equation, via a conformal mapping combined with a
transformation $\psi\mapsto\psi+f(z)+\overline{f(z)}$. In other words,
the explicit $z$-dependence in (\ref{psi-eqn}) could then be removed,
at the cost of more complicated global conditions on the field $\psi$.


\section{Symmetries and the $n=1$ solution}
We are taking $H(z)=\det\Phi$ to be a polynomial of degree~$n$,
so $H(z)=p_0z^n+\ldots+p_n$, where the $p_k$ are complex constants.
The Hitchin equations are conformally invariant, but the quantity
$|p_0|$ (say) sets the length scale, and from now on we shall fix this scale
by setting $|p_0|=1$.
The equations also have a rotational symmetry
$z\mapsto\ee^{\ii\chi}z$, as well as a global phase symmetry
$\Phi\mapsto\ee^{\ii\phi}\Phi$, and these are in effect linked by the
argument of $p_0$; we fix this phase by setting $p_0=1$.

Some solutions are rotationally-symmetric, which clearly can only
happen if $H(z)=z^n$. We say that the field $(\Phi,A_{\zb})$
with $\det\Phi=z^n$ is rotationally-symmetric if the rotated field
\begin{equation}  \label{rotsym}
  \widetilde{\Phi}(z)={\rm e}^{-\ii n\chi/2}\Phi({\rm e}^{\ii\chi}z),
  \quad  \widetilde{A}_{\zb}(z)={\rm e}^{-\ii\chi}A_{\zb}({\rm e}^{\ii\chi}z)
\end{equation}
is gauge-equivalent to $(\Phi(z), A_{\zb}(z))$.
Rotationally-symmetric solutions were studied a long time ago, in
effect only for the case $n=0$ \cite{S81, S84}; and also more recently,
where it was shown that in certain cases the solutions can be expressed in
terms of Painlev\'e transcendents \cite{MW93, GMN13}.

Although most of the global phase symmetry has been removed by fixing the
phase of $p_0$ as we did above, there is a remnant left over, namely
$\Gamma:\Phi\mapsto-\Phi$. Now $\Gamma$ changes the sign of
the function $\alpha$ appearing in the gauge potential (\ref{A-form}),
so the $\Gamma$-invariant fields are precisely those with $\alpha=0$.
By putting $\chi=2\pi$ in (\ref{rotsym}), we see that a rotationally-symmetric
field with $n$ odd is also $\Gamma$-invariant.
But for $n$ even, there are rotationally-symmetric fields which are not
$\Gamma$-invariant, as we shall see in the next section.

For the remainder of this section, let us focus on the $n=1$ case.
We can use translation freedom to set $H(z)=z$. There is a
rotationally-symmetric $n=1$ solution which has its gauge field $F$
peaked at $z=0$; it is given by (\ref{psi-form}) and (\ref{A-form}) with
$\mu_+=z$, $\mu_-=-1$, $\psi=\psi(r)$ and $\alpha=0$, where $r=|z|$.
If we change variables to $t=r^{3/2}$ and $h(t)=\ee^{-\psi/2}t^{-1/3}$, then
the remaining field equation (\ref{psi-eqn}) becomes a particular case of
the Painlev\'e-III equation, namely
\begin{equation}  \label{PIII_1}
  h''-\frac{(h')^2}{h}+\frac{h'}{t}+\frac{4}{9h}-\frac{4h^3}{9}=0.
\end{equation}
Figure~1 shows the magnitude $|F(r)|$ of the corresponding gauge field,
obtained by solving (\ref{PIII_1}) numerically. We see that
the gauge field is peaked at $z=0$, and has a radius of order unity;
it is in effect abelian, and so we can use Green's theorem to compute
$\int |F|\,dx\,dy = \pi/2$.
\begin{figure}[htb]
\begin{center}
\includegraphics[scale=0.5]{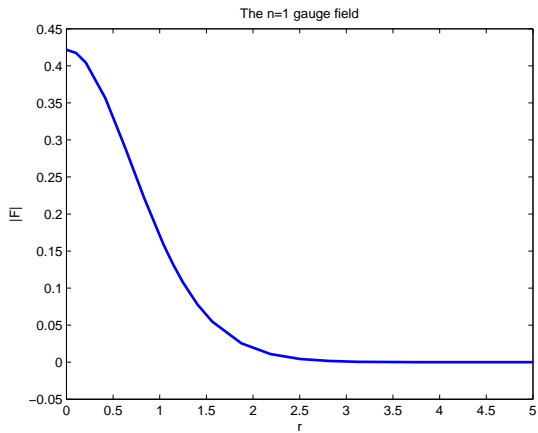}
\caption{The profile function $w(r)$ and the gauge field $|F(r)|$
for the $n=1$ solution. \label{fig1}}
\end{center}
\end{figure}

Outside the core region around $z=0$, there is an approximation
to the field, or `limiting configuration' \cite{MSWW14, MSWW15},
which turns out to be a useful one: namely
\begin{equation}  \label{1-lump_approx}
  \Phi=\sqrt{z}\,\ii\sigma_1,  \quad A_x=0,  \quad A_y=0  \mbox{ for $x>0$},
    \quad A_y=\frac{\pi}{2}\delta(y)\,\ii\sigma_3 \mbox{ for $x<0$}.
\end{equation}
Here the square root is branched along the negative $x$-axis, where the
gauge potential has a delta-function singularity. The corresponding gauge
field $F$ is supported at $z=0$: in fact, it is a 2-dimensional
delta-function $F=-(\pi/2)\,\delta^2(x,y)\,\ii\sigma_3$. The branch discontinuity
is gauge-removable, but the singularity at $z=0$ is not. For a given $\Phi$
as in (\ref{1-lump_approx}), the gauge field could instead be proportional
to $\cos(\omega)\sigma_2+\sin(\omega)\sigma_3$, so there is
an angle $\omega$ associated with the lump. In the $n=1$ case, this angle
is gauge-removable; but for $n>1$ the relative phase between lumps
plays a role, as shall see below.


\section{The $n=2$ case}
The general form of $H(z)$ is $H(z)=z^n+p_1z^{n-1}+\ldots+p_n$, and
the singular approximation generalizing (\ref{1-lump_approx}),
namely $\Phi=\sqrt{H}\,\ii\sigma_1$, can be used
to determine which of the coefficients $p_k$ are moduli.
The convergence of (\ref{L2norm}) is determined by the asymptotic
behaviour of the fields as $r\to\infty$, and the singular approximation is
a good one in this asymptotic region.
The $L^2$ norm of the derivative $\pa\Phi/\pa p_k$, for the singular field, is
\begin{equation}  \label{NormPhidotsq}
  \left\|\frac{\pa\Phi}{\pa p_k}\right\|^2 = 
      \int_{\RR^2} \, \frac{|z|^{2n-2k}}{4|H|}\,dx\,dy.
\end{equation}
Assuming that the zeros of $H$ are simple, this integral converges
if and only if $k$ lies in the range $(n+3)/2\leq k\leq n$. So for these values
of $k$, the coefficient $p_k$ is a complex modulus.
In particular, the $n=1$ and $n=2$ cases do not admit moduli, whereas
for $n\geq3$ we do get moduli.

In addition to the coefficients $p_k$, there are other moduli: these are the
relative phases between the $n$~lumps.
If~$n$ is odd, then $(n-1)/2$ of the coefficients $p_k$ are complex moduli,
which together with $n-1$ relative phases gives a total of $2(n-1)$
real moduli. If~$n$ is even, then $(n-2)/2$ of the coefficients are complex
moduli. In this case, there are $n-1$ relative phases, but one combination
of them does not have $L^2$ variation, so we end up with a total of
$2(n-2)$ real moduli in the even case.
In particular, the $n=3$ and $n=4$ cases each have a 4-dimensional
moduli space $\cM$.

The $\Gamma$-invariant fields are exactly those for which the lumps are
parallel or antiparallel, in other words the relative phases are $0$ or $\pi$.
This may be understood in the gauge (\ref{psi-form}, \ref{A-form}),
as follows. Let $\{z_1,\ldots,z_n\}$ denote the roots of $H(z)$.
Then the relative phase between the lump at $z_j$ and the one at $z_k$
is the angle in ${\mathfrak su}(2)$ between the
gauge fields $F^{(j)}$ and $F^{(k)}$ at these two points.
To make sense of this, we have to specify how to compare the isovectors
$F$ at the two points. But for $\Gamma$-invariant fields we have
$\alpha=0$ in (\ref{A-form}), and then 
parallel-propagation between the two points does not change $F$ at
all; so we conclude that $F^{(j)}$ and $F^{(k)}$, both being in the
$\sigma_3$-direction, are either parallel or antiparallel.

We now examine some features of the $n=2$ case. Let ${\cal S}$
denote the set of solutions which are invariant under $\Gamma$,
up to gauge. Then $\cS$ has two components, namely $\cS_+$
where $F^{(1)}$ and $F^{(2)}$ are parallel, and $\cS_-$ where they
are antiparallel. Let $\cR$ denote the space of rotationally-symmetric
fields. Then $\cS_-\cap\cR$ has the two gauge fields cancelling: indeed,
it is the explicit degenerate solution
\begin{equation}  \label{Explicit2}
  \Phi=z\ii\sigma_1, \quad A_{\zb}=0.
\end{equation}
The set $\cR$ of rotationally-symmetric solutions
is a 1-parameter family of fields interpolating between
(\ref{Explicit2}) and the non-explicit field $\cS_+\cap\cR$. This family
can be obtained as solutions of a boundary-value problem for a single
real-valued function, as follows.

We work in the gauge (\ref{psi-form}, \ref{A-form}), with $\mu_+=z^2$
and $\mu_-=-1$. In the $n=2$ case, the rotational-symmetry condition
becomes $\psi=\psi(r)$ and $\alpha=\alpha(r)$.
Then the general solution of equation (\ref{alpha-eqn}) is
$\alpha=B_+/M_++\ii B_-/M_-$, where $B_{\pm}$ are real constants and
\[
   M_{\pm}=r^2\ee^{\psi/2}\mp\ee^{-\psi/2}.
\]
Now the boundary condition $F\to0$ as $r\to\infty$ implies that $B_+=0$,
so we end up with an expression for $\alpha$ in terms of $\psi(r)$ and the
real constant $B=B_-$, namely $\alpha=\ii B/M_-$. This can then
be substituted into (\ref{psi-eqn}) to give an equation for $\psi(r)$, namely
\begin{equation}  \label{rotsym2}
  \psi'' + r^{-1}\psi' = 2[1+4B^2(r^2\ee^{\psi/2}+\ee^{-\psi/2})^{-2}]
    (r^4\ee^{\psi}-\ee^{-\psi}).
\end{equation}
The boundary conditions on $\psi$ are $\psi'(0)=0$ and
$\psi(r)\sim-2\log(r)$ as $r\to\infty$. 
In the $B=0$ case (where $\alpha=0$), it is again a Painlev\'e-III equation,
but with a different parameter. Its nature for general $B$ might be worth
investigating, for example via a Painlev\'e analysis.
\begin{figure}[htb]
\begin{center}
\includegraphics[scale=0.5]{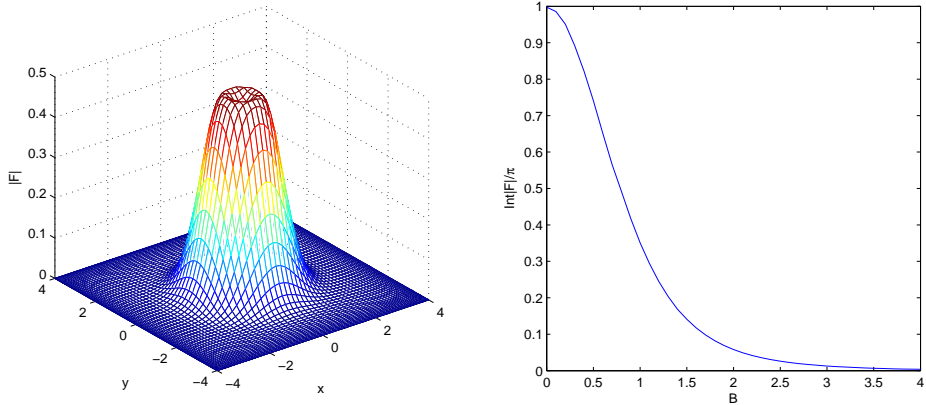}
\caption{The gauge field $|F|$ for a rotationally-symmetric $n=2$ field
with $B=0$; and $\pi^{-1}\int |F|\,d^2x$ as a function of $B$. \label{fig2}}
\end{center}
\end{figure}

For each value of $B\geq0$, we get a rotationally-symmetric solution,
and this family is illustrated in Figure~2.  The left-hand plot is the $B=0$ case,
and shows the norm $|F|$ of the gauge field as a
function of $x$ and $y$; this was obtained by solving (\ref{psi-eqn}) numerically
with $\alpha=0$. The field $|F|$ is peaked on a circle centred at
$z=0$, and $\int |F|\,d^2x=\pi$.
The right-hand diagram shows the result of solving the
boundary-value problem (\ref{rotsym2}) numerically, and plots
$\pi^{-1}\int |F|\,d^2x$ as a function of $B$. The limit $B\to\infty$ is
the degenerate solution (\ref{Explicit2}) where $F=0$, whereas $B=0$ is the
solution depicted in the left-hand plot.


\section{The $n=3$ case}
If $n=3$, then $H(z)$ has the form $H(z)=z^3+p_1z^2+p_2z+p_3$.
The constants $p_1$ and $p_2$ are complex parameters, whereas
$p_3$ is a complex modulus. By translation of $z$, we may set $p_1=0$.
We can also use a rotation to make $p_2$ real. This leaves us with
\begin{equation}  \label{H3poly}
   H(z)=z^3+az-K,
\end{equation}
where $a>0$ is a fixed parameter, and $K$ is a complex modulus.
For each value of $a$, we have a 4-dimensional moduli space $\cM_a$,
on which the local coordinates are the real and imaginary parts of $K$
together with two relative phases $\eta_1$ and $\eta_2$.
Note that $\cM_0$ admits a rotational Killing vector, since
$K\mapsto\ee^{\ii\nu}K$ is then an isometry.

The asymptotic region of $\cM_a$ is where $|K|\gg1$, and in this region
the singular approximation $\Phi=\sqrt{H}\,\ii\sigma_1$ is a good one
\cite{MSWW14, MSWW15}.
The reason for this is that the lumps become smaller as they separate,
{\it ie.}\ as $K\to\infty$. For example, if $H(z)=z^3-b^3$,
then $H(z)\approx3b^2(z-b)$ near $z=b$, and so the lump at
$z=b$ has characteristic size $1/(3b^2)$, and in effect it approaches
a delta-function as $b\to\infty$. So the singular approximation
is an accurate one in the asymptotic region of $\cM$ where the zeros
of $H(z)$ are well-separated.

Let $\{z_1,z_2,z_3\}$ denote the zeros of (\ref{H3poly}).
The gauge field near $z_j$ has the approximate form
\[
  F^{(j)} = \frac{\ii\pi}{2}[\cos(\omega_j)\,\sigma_2 + 
    \sin(\omega_j)\,\sigma_3]\,\delta(z-z_j).
\]
This gives us three angles $\{\omega_1,\omega_2,\omega_3\}$ in
the $\sigma_2\sigma_3$-plane, and hence two relative angles, say
$\eta_1=\omega_1-\omega_3$ and $\eta_2=\omega_2-\omega_3$.
The asymptotic moduli are $K$, $\eta_1$ and $\eta_2$, and
the structure is that of a $T^2$-bundle over the $K$-space, with
$\eta_1,\eta_2\in[-\pi,\pi]$ being the $T^2$ coordinates. This bundle is
twisted: if the phase of $K$ goes from $0$ to $2\pi$, thereby permuting
the zeros of $H(z)$, then the angles $(\eta_1,\eta_2)$ are transformed
by an element $\Upsilon$ of the modular group SL(2,$\ZZ$), namely
\[
   \Upsilon:(\eta_1,\eta_2)\mapsto(-\eta_2,\eta_1-\eta_2).
\]
In fact, $\Upsilon$ is a cube root of unity in SL(2,$\ZZ$).

One may now calculate the asymptotic metric on the moduli space
from the integral formula (\ref{L2norm}), by using the singular approximation.
The calculation is analogous to that described in \cite{MW13}, and will
only be sketched here. For large $|K|$, the parameter $a$ may be ignored,
and so the singular approximation is $\Phi=\sqrt{z^3-K}\,\ii\sigma_1$.
Differentiating this with respect to $K$ gives a vector $V_1$, and its norm
is $\|V_1\|^2=\int|d\sqrt{z^3-K}/dK|^2\,|dz|^2$. The structure of the
perturbation equations (\ref{PertEqns}) enables us to write down three
other vectors $\{V_2,V_3,V_4\}$, giving an orthogonal tetrad $\{V_{\mu}\}$
of vectors all having the same norm. For each of the $V_{\mu}$, we can
compute the corresponding change in the coordinates $K,\eta_1,\eta_2$.
This then gives us a coordinate expression for the metric, which turns
out to be
\begin{equation}  \label{SingMetric}
   ds^2 = c |K|^{-1/3} |dK|^2 + \frac{1}{\sqrt{3}}
        \left(d\eta_1^2+d\eta_2^2-d\eta_1\,d\eta_2\right),
\end{equation}
where
\[
  c = \frac{3\sqrt{3}}{4}\left(\int_0^1\frac{du}{\sqrt{1-u^3}}\right)^2
     \approx 2.554.
\]
This metric is flat, with a conical singularity at $K=0$. It is of type
ALG \cite{CK02}, since the 4-volume of a ball with geodesic radius
$R=|K|^{5/6}$ is proportional to $R^2$ for large $R$. It is invariant
under the modular transformation $\Upsilon$, and therefore well-defined.

Now $\Gamma:\cM_a\to\cM_a$ is an isometry, and so
the subset $\cS\subset\cM_a$ of $\Gamma$-invariant fields is a
2-dimensional totally-geodesic submanifold of $\cM_a$. In what follows,
we study the geometry of this surface $\cS$. Now $\cS$ has two
components: one of them is $\cS_+$, where the three lumps are parallel
($\eta_1=\eta_2=0$); and the other is $\cS_-$, where two lumps are
parallel and the third is antiparallel to them. Let us first consider
$\cS_+$, which is diffeomorphic to $\RR^2$ and has $K$ as a global
complex coordinate. The induced metric on $\cS_+$ has the form
$ds^2=\Omega(K,\Kb)\,|dK|^2$. From (\ref{SingMetric}) we expect that
$\Omega\sim c\,|K|^{-1/3}$ as $|K|\to\infty$. In fact,
$ds^2= c\,|K|^{-1/3}\,|dK|^2$ is a cone, and so
$\cS_+$ is this cone with its vertex smoothed.

A numerical calculation can be used to obtain information about
the geometry of $\cS_+$ in its central region. The procedure is as follows:
solve (\ref{psi-eqn}) with $\alpha=0$ numerically, for a range of
values of $K$; obtain tangent vectors $V=(\dot{\Phi},\dot{A}_{\zb})$
by evaluating the difference in the fields for neighbouring values of
$K$ and then projecting orthogonal to the gauge orbits; evaluate the integral
(\ref{L2norm}) to calculate $\Omega$; and finally compute
the Gaussian curvature $C$ from $\Omega$. This numerical procedure
gives results that are consistent with the expected asymptotic form:
for example, in the case $a=0$
one finds $1-\Omega|K|^{1/3}/c\approx10^{-3}$ when $|K|=3$.
The results are illustrated in Figure~3, which plots the curvature
$C$ versus $K$, for two values of the parameter $a$.
\begin{figure}[htb]
\begin{center}
\includegraphics[scale=0.5]{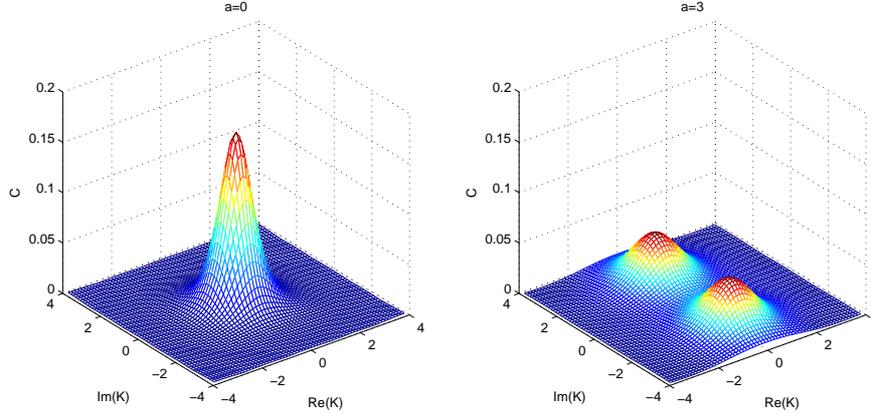}
\caption{Gaussian curvature $C$ of $\cS_+$, for $a=0$ and $a=3$.\label{fig3}}
\end{center}
\end{figure}
The left-hand plot is for $a=0$, and is rotationally-symmetric as expected.
The right-hand one is for $a=3$, and we see that the curvature is 
peaked at two points: these are the values $K=\pm2(-a/3)^{3/2}$
for which the polynomial $H(z)$ has a double root, which means that
two of the three lumps coincide.

\begin{figure}[htb]
\begin{center}
\includegraphics[scale=0.5]{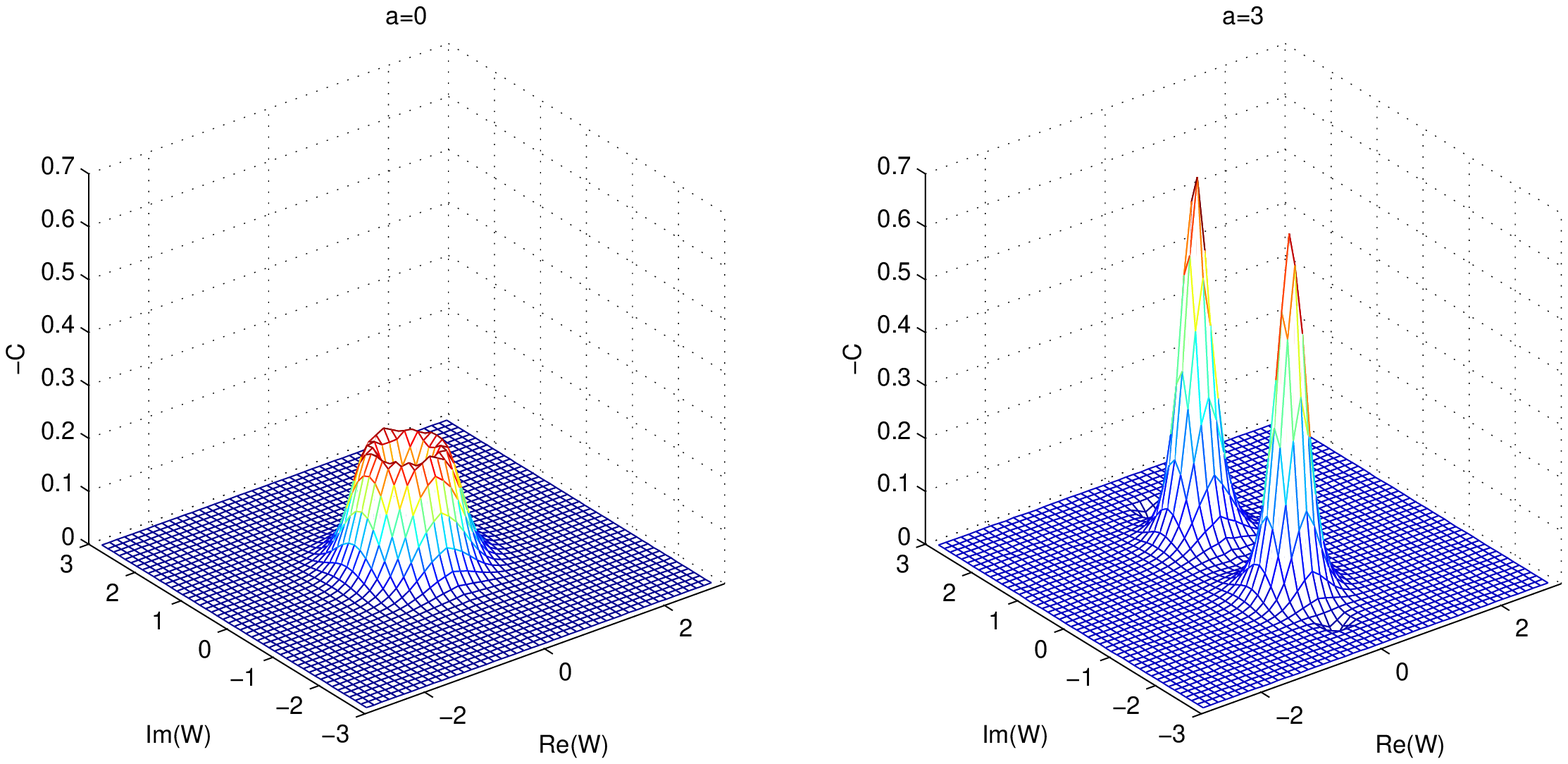}
\caption{Curvature of $\cS_-$, for $a=0$ and $a=3$.\label{fig4}}
\end{center}
\end{figure}
Next we do the same thing for the other component $\cS_-$.
The surface $\cS_-$ is a triple cover of the $K$-plane, since any one
of the three lumps can be the odd one out (antiparallel to the other two).
Indeed, $\cS_-$ has a global complex coordinate $W$, where
$W^3+aW-K=0$. Repeating the procedure described above leads to
Figure~4. In this case, the Gaussian curvaure $C$ is
(predominantly) negative, and $-C$ is plotted. Once again
the $a=0$ case is rotationally-symmetric, while in the $a=3$ case
the curvature has four peaks. There are two values of $W$, and hence
$K$, for which a lump concides with the antilump, leaving only a single
lump in $\RR^2$; the curvature is negative at these values of $W$
(in the right-hand plot of Figure~4, this is at $W=\pm\ii$).
And there are two values of $W$
for which the two parallel lumps coincide; the curvature is
positive at those values ($W=\pm2\ii$ in Figure~4).

The picture therefore is that the moduli space $\cM_a$ has an
asymptotically-conical geometry; the vertices of this cone, which are
located at the values of $K$ for which $H(z)=\det\Phi$ has coincident zeros,
are smoothed out, and the curvature of $\cM_a$ is concentrated at these loci.

\section{Remarks}

The natural energy functional for the Hitchin system, arising as a
dimensional reduction of the Yang-Mills action in $\RR^4$, is
\begin{equation}  \label{Energy}
E=\int_{\RR^2}\left(|D_j\Phi|^2+|F|^2+\quar|[\Phi,\Phi^*]|^2\right)\,dx^1\,dx^2.
\end{equation}
Solutions of the Hitchin equations (\ref{H1}) are critical popints of
(\ref{Energy}). However, for the soliton-like solutions described in this
paper, the integral (\ref{Energy}) diverges, on account of the
$|D_j\Phi|^2$ term. One could try to regularize the integral by
subtracting a fixed function from the integrand, in other words one
that does not depend on the moduli. But a calculation with the
approximate solution quickly shows that this does not work:
a moduli-dependent part remains which is at least logarithmically
divergent. So it remains unclear as to whether the
energy (\ref{Energy}) can be regularized in a satisfactory way.

The information which determines a solution may also be understood
in terms of spectral data, consisting of a bundle over a spectral curve,
which is the zero-set of $F(z,t)=\det\left(\Phi(z)-t\right)$. In the SU(2)
case with $n=3$ described above, the spectral curve is the standard elliptic
curve $t^2=z^3+az-K$.
The spectral data are preserved under the generalized Nahm transform.
The details of this transform depend on the boundary conditions being
imposed, and several different examples of Nahm transforms applied to
the Hitchin equations have been studied \cite{CK01, B06, S07, FJ08}.

In this paper, we have only considered one particular moduli space,
namely that of smooth $n=3$ solutions with gauge group SU(2).
The same sort of methods can readily be applied to other cases,
where $n$ and the gauge group are different and where one allows
singularities. One would expect there to be isometries between some
of the resulting moduli spaces; a first step would be to classify these
spaces up to isometry, using the Nahm transform, along the lines of
what was done for doubly-periodic monopoles \cite{CW12, C14}.

\vspace{0.5cm}\noindent{\bf Acknowledgment.}
I am grateful to Maciej Dunajski for helpful comments.
This work was supported by the UK Particle Science and Technology
Facilities Council, through the Consolidated Grant ST/L000407/1.


\end{document}